\documentclass[apj]{emulateapj}
\usepackage{color}

\newcommand{\etal}{et~al.}

\newcommand{\ergscm}{erg~s$^{-1}$~cm$^{-2}$}
\def\gsim{\mathrel{\rlap{\lower 4pt \hbox{\hskip 1pt $\sim$}}\raise 1pt \hbox {$>$}}} 
\def\lsim{\mathrel{\rlap{\lower 4pt \hbox{\hskip 1pt $\sim$}}\raise 1pt \hbox {$<$}}} 

\shorttitle{Extremely high polarization in 2010 outburst of 3C~454.3} \shortauthors{Sasada et~al.}

\begin{document}

\title{Extremely high polarization in 2010 outburst of blazar 3C~454.3}

\author{Mahito \textsc{Sasada}\altaffilmark{1},
        Makoto \textsc{Uemura}\altaffilmark{2},
        Yasushi \textsc{Fukazawa}\altaffilmark{3},
        Hajimu \textsc{Yasuda}\altaffilmark{3}, 
        Ryosuke \textsc{Itoh}\altaffilmark{3},
        Kiyoshi \textsc{Sakimoto}\altaffilmark{3}, \\
        Yuki \textsc{Ikejiri}\altaffilmark{3},
        Michitoshi \textsc{Yoshida}\altaffilmark{2}, 
        Koji S. \textsc{Kawabata}\altaffilmark{2},
        Hiroshi \textsc{Akitaya}\altaffilmark{2},
        Takashi \textsc{Ohsugi}\altaffilmark{3}, 
        Masayuki \textsc{Yamanaka}\altaffilmark{4}, 
        Tomoyuki \textsc{Komatsu}\altaffilmark{3}, 
        Hisashi \textsc{Miyamoto}\altaffilmark{3},
        Osamu \textsc{Nagae}\altaffilmark{3},
        Hidehiko \textsc{Nakaya}\altaffilmark{5},
        Hiroyuki \textsc{Tanaka}\altaffilmark{3},
        Shuji \textsc{Sato}\altaffilmark{6}, \\ and
        Masaru \textsc{Kino}\altaffilmark{1}
}

\altaffiltext{1}{Department of Astronomy, Kyoto University, Kitashirakawa-Oiwake-cho, Sakyo-ku, Kyoto 606-8502, Japan; sasada@kusastro.kyoto-u.ac.jp} 
\altaffiltext{2}{Hiroshima Astrophysical Science Center, Hiroshima University, Higashi-Hiroshima, Hiroshima 739-8526, Japan} 
\altaffiltext{3}{Department of Physical Science, Hiroshima University, Kagamiyama 1-3-1, Higashi-Hiroshima 739-8526, Japan} 
\altaffiltext{4}{Kwasan Observatory, Kyoto University, 17-1 Kitakazan-ohmine-cho, Yamashina-ku, Kyoto, 607-8471, Japan} 
\altaffiltext{5}{National Astronomical Observatory of Japan, Osawa, Mitaka, Tokyo 181-8588, Japan}
\altaffiltext{6}{Department of Physics, Nagoya University, Furo-cho,
Chikusa-ku, Nagoya 464-8602, Japan}


\begin{abstract}

 The gamma-ray-detected blazar 3C~454.3 exhibits dramatic flux and
 polarization variations in the optical and near-infrared bands. In
 December 2010, the object emitted a very bright outburst. We monitored
 it for approximately four years (including the 2010 outburst) by
 optical and near-infrared photopolarimetry. During the 2010 outburst,
 the object emitted two rapid, redder brightenings, at which the
 polarization degrees (PDs) in both bands increased significantly and
 the bands exhibited a frequency-dependent polarization. The observed
 frequency-dependent polarization leads us to propose that the
 polarization vector is composed of two vectors. Therefore, we separate
 the observed polarization vectors into short and long-term components
 that we attribute to the emissions of the rapid brightenings and the
 outburst that varied the timescale of days and months,
 respectively. The estimated PD of the short-term component is greater
 than the maximum observed PD and is close to the theoretical maximum
 PD. We constrain the bulk Lorentz factors and inclination angles
 between the jet axis and the line of sight from the estimated PDs. In
 this case, the inclination angle of the emitting region of short-term
 component from the first rapid brightening should be equal to
 90$^{\circ}$, because the estimated PD of the short-term component was
 approximately equal to the theoretical maximum PD. Thus, the Doppler
 factor at the emitting region of the first rapid brightening should be
 equal to the bulk Lorentz factor.

\end{abstract}


\section{Introduction}

Blazars are a type of active galactic nuclei with relativistic
jets that are widely believed to be viewed at small angles to the line
of sight. Blazars frequently show violent variations in flux and
polarization, which vary on timescales ranging from minutes
\citep{Aharonian07,Sasada08} to years \citep{Sillanpaa96}. On the
timescale of months, blazars may turn out to be over 10 times brighter
than in their quiescent state. These brightening phenomena are often
called ``outburst''.

The blazar 3C~454.3 with a redshift $z=0.859$ \citep{Jackson91} is one
of the most famous blazar because it has emitted several large-amplitude
outbursts. In 2005, it emitted a dramatic outburst that covered the
range from radio to hard X-ray bands
\citep{Fuhrmann06,Pian06,Giommi06,Villata07}. The maximum brightness in
the optical band reached $R=12.0$. Many authors reported that prominent
outbursts occurred from 3C~454.3 in 2007, 2008 and 2009 and covered the
range from radio to gamma-ray bands. Gamma-ray emission from the 2007
outburst was detected by the AGILE satellite
\citep{Vercellone08,Donnarumma09,Vercellone10}, and brightenings in
other wavelengths were also detected
\citep{Villata08,Raiteri08a,Raiteri08b,Villata09a,Sasada10}. 
\citet{Ghisellini07} and \citet{Sikora08} proposed two possible
explanations for the origin of seed photons for inverse Compton
scattering emission in the GeV gamma-ray band. A large-amplitude
gamma-ray brightening in the 2008 outburst was detected by the {\it
Fermi Gamma-ray Space Telescope} \citep{Bonning09a,Abdo09}. Based on
intensive monitoring, it was determined that the flux variations in the
optical bands lagged no more than one day behind those of the GeV
gamma-ray band
\citep{Vercellone09a,Donnarumma09,Bonning09a,Vercellone10}. In 2009, 
an outburst from 3C~454.3 occurred over all wavelengths: in the
gamma-ray band it was detected by {\it Fermi} and AGILE
\citep{Striani09a,Striani09b,Escande09,Striani10,Pacciani10,Ackermann10},
in the X-ray band it was detected by {\it INTEGRAL}
\citep{Vercellone09b}, {\it Swift}/XRT \citep{Sakamoto09}, and 
{\it Swift}/BAT \citep{Krimm09}, in the optical bands it was detected by
many groups \citep{Villata09b,Bonning09b,Sasada09,Raiteri11,Sasada12},
and in the radio bands it was detected by \citet{Raiteri11} and 
\citet{Jorstad13}.

In September 2010, a prominent outburst from 3C~454.3 was detected in
the GeV gamma-ray band by {\it Fermi} and AGILE
\citep{Abdo11,Vercellone11}. The peak flux reached 85$\times$10$^{-6}$
photons~cm$^{-2}$~s$^{-1}$ in December 2010. The optical continuum and
emission-line fluxes increased simultaneously in this outburst
\citep{Vercellone11,Leon-Tavares13}, and the X-ray and radio fluxes also 
increased \citep{Wehrle12,Jorstad13}.

In the present study, we report the results of using the Kanata
telescope to monitor emission from 3C~454.3 using optical and near-infrared
(NIR) photopolarimetry from July 2007 to January 2011: a period that
includes the 2010 outburst. In December 2010, the polarization degree
(PD) increased in the rapid brightenings of the 2010 outburst and the
polarization difference between the optical and NIR bands became
frequency dependent (this phenomenon is called
``frequency-dependent polarization'' or FDP). The temporal variations of
polarization angle (PA)
indicate that no rotation event occurred during this outburst. The paper
is organized as follows: In section~2, we present the methods used for
observation and analysis. In section~3, we report the results of the
photometric and polarimetry observation. In section 4, we discuss 
the origin of the observed FDP and emission regions of the observed
rapid brightenings and outburst. The conclusion is given in section 5.

\section{Observations}

We monitored 3C~454.3 by multi-band photopolarimetry from July 2007 to
January 2011 (JD~2454618---2455590). The monitoring was performed in the
optical and NIR bands by attaching TRISPEC \citep{Watanabe05} and HOWPol
\citep{Kawabata08} to the 1.5-m Kanata telescope at the
Higashi-Hiroshima Observatory. TRISPEC enables simultaneous
photopolarimetric observations in the optical band and two NIR
bands. HOWPol can also make photopolarimetric measurements in the
optical band. Unfortunately during the entire monitoring period, one
channel (IR-2) of the two NIR arrays and an optical CCD of the TRISPEC
were at times not available because of readout errors. We compensated
for the lack of optical data by using the HOWPol. Thus, we measured the
$V$- and $J$-band photopolarimetry at the same time. An observation
sequence unit consisted of successive exposures with a half-wave plate
set at four different angles. The polarization for the given observation
sequence unit was derived from the set of four exposures.

We adjusted the integration time depending on the sky conditions and
the brightness of the object. Typical integration times were 170~s in
both bands. For all images, the bias and dark were subtracted and the
image field was flattened before performing the aperture photometry. By
comparing with star~13 listed in \citet{Gonzalez01}, which appeared in
the same frame as 3C~454.3, we were able to conduct differential
photometry. For star~13, we used the magnitudes $V=13.587$ and
$J=11.858$ \citep{Gonzalez01,Skrutskie06}, and (for 3C~454.3) we adopted
a Galactic extinction of 0.349 and 0.097~mag in the $V$ and $J$ bands,
respectively \citep{Schlegel98}. We converted the observed magnitudes
into fluxes by referring to \citet{Fukugita95} and \citet{Bessell98}.

We confirmed that the instrumental polarization of TRISPEC was less
than 0.1~\% in the $V$ and $J$ bands by using the unpolarized standard
stars. Thus, we did not apply any correction for the instrumental
polarization. However, we applied the instrumental depolarization
factors of 0.827 and 0.928 in the $V$ and $J$ bands, respectively. The
zero point of the PA was defined by the standard system (measured from
north to east) by observing the polarized standard stars: HD~19820 and
HD~25443 \citep{Wolff96}.

A large instrumental polarization ($\sim$4\%) appears from HOWPol
because of the asymmetric reflection of the incident light from the
tertiary mirror of the Kanata telescope. The instrumental polarization
can be expressed as a function of a declination of 3C~454.3 and an hour
angle at the observation. We calculate the intrinsic polarization by
subtracting the instrumental polarization from the raw polarization. The
deviation of the instrumental polarization is estimated by observing the
unpolarized standard stars. The error from instrumental polarization
is less than 0.5$~{\%}$, which we calibrate from the deviation.

\section{Results}


\begin{figure}
\begin{center}
\begin{tabular}{c}
\includegraphics[scale=1.1]{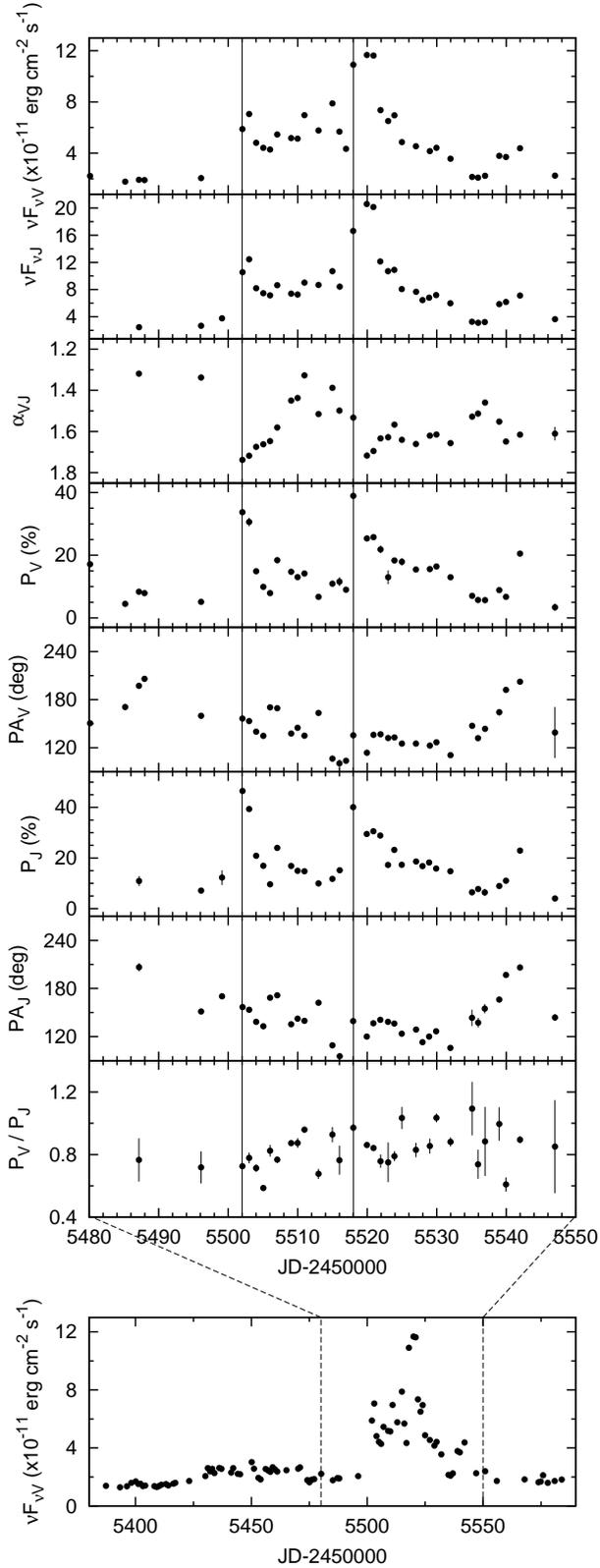}
\end{tabular}
  \caption{Variations of observational parameters of 3C~454.3 during
 2010 outburst. From top to bottom, we show the light curves in the $V$
 and $J$ bands (units are {\ergscm}), temporal variations of the
 spectral index between the $V$ and $J$ bands, $\alpha_{VJ}$, PDs and
 PAs (units are \% and degrees, respectively) in the $V$ and $J$ bands,
 and the ratio $P_{V}/P_{J}$ between the PDs in both bands in the 2010
 outburst. Solid lines show the first and second rapid brightenings in
 JD~2455502 and 2455518. The bottom panel shows the $V$-band light curve
 during the entire period of 2010 outburst.}
\label{lc}
\end{center} 
\end{figure}

\begin{figure*}
\begin{center}
\begin{tabular}{c}
\includegraphics[scale=1.15]{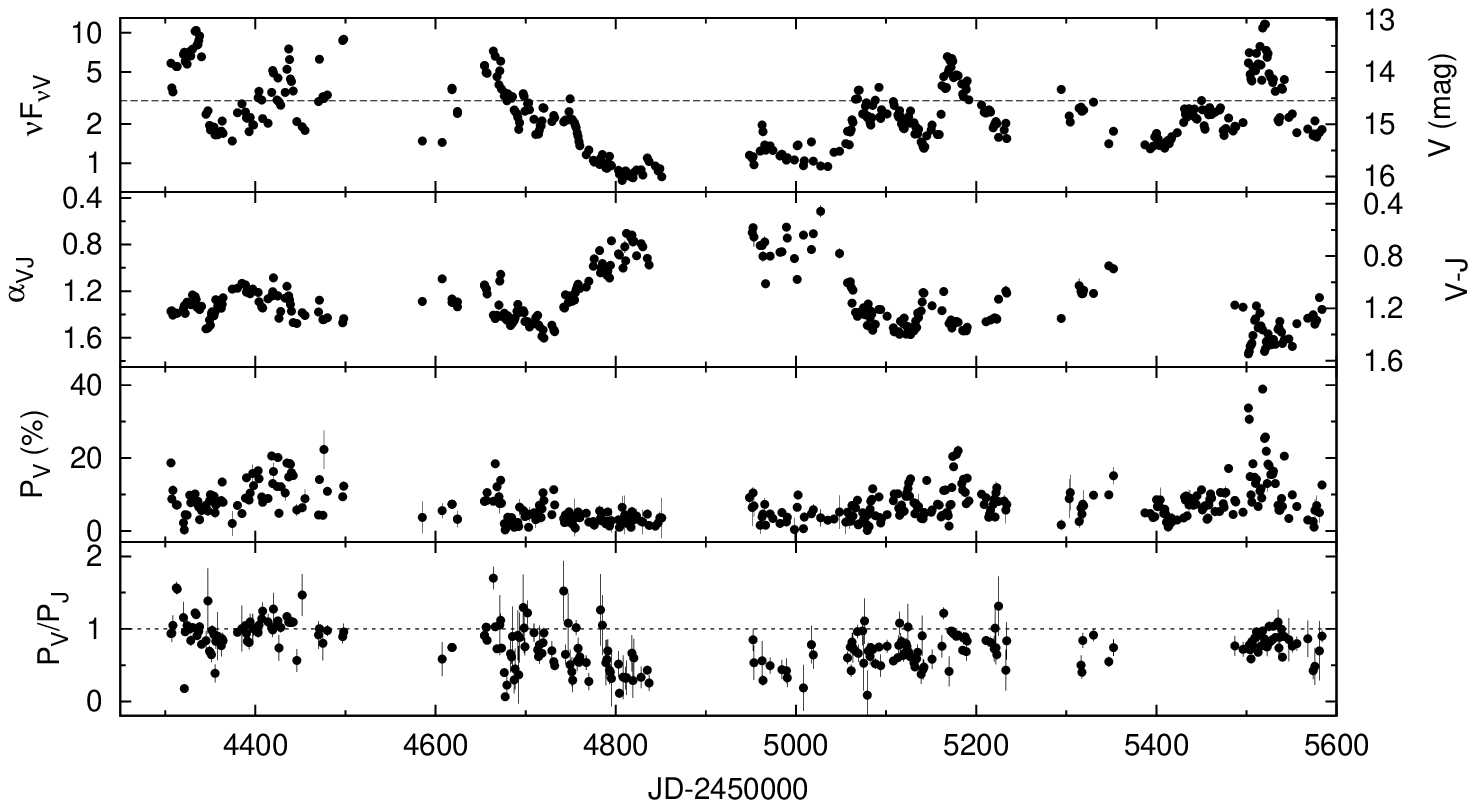}
\end{tabular}
  \caption{Four-year variations of 3C~454.3 from July 2007 to January
 2011. From top to bottom, the panels show the $V$-band light curve
 plotted on a logarithmic scale (same unit as for Figure~\ref{lc}),
 the temporal variations of $\alpha_{VJ}$, the PD in the $V$ band (in
 percent), and the ratio $P_{V}/P_{J}$ between the PDs in the $V$ and
 $J$ bands, respectively. In the $V$-band light curve, we also show an
 upper limit imposed on the observed flux by the thermal radiation from
 the accretion disk. See the text for details.}
\label{wh}
\end{center} 
\end{figure*}

Figure~\ref{lc} shows the temporal variations of the observational
parameters during our 2010 monitoring. The bottom panel of
Figure~\ref{lc} shows the $V$-band light curve during the entire 2010
monitoring period. From top to bottom, the panels show the $V$- and
$J$-band light curves, the temporal variations of spectral
index $\alpha_{VJ}$  ($F_{\nu}\;{\propto}\;{\nu}^{-\alpha}$), the PDs
and PAs in the $V$ and $J$ bands, and the ratio $P_{V}/P_{J}$ between
the PDs in both bands during the 2010 outbursts. The units of flux, PD
and PA are {\ergscm}, percent and degrees, respectively. During the 2010
outburst, there were two rapid brightenings at JD~2455502 and 2455518
shown by solid lines in Figure~\ref{lc}. The maximum fluxes in the $V$
and $J$ bands were 1.167$\times$10$^{-10}$ and 2.060$\times$10$^{-10}$
{\ergscm}, respectively, in JD~2455520. After the peak, the object
gradually became faint. The object became redder as it brightened, which
we call the ``redder-when-brighter'' phenomenon. In general, blazars
become bluer as they brighten, (``bluer-when-brighter'';
\citealt{Racine70}). However, in previous publications, 3C~454.3 was
reported to be redder-when-brighter
\citep{Miller81,Villata06,Raiteri07,Sasada10}.

The polarization vectors in both bands were variable and exhibited
almost similar behavior. In the $V$ and $J$ bands, the PDs reached as
high as 38.9$\pm$0.4 and 46.5$\pm$2.1~\%, respectively, during the rapid
brightenings. Although the PD characteristics were approximately the
same, the $J$-band PD increased more than the $V$-band PD. Thus,
3C~454.3 exhibits a FDP, which means the PDs differ between two selected
wavelengths. The eighth panel of Figure~\ref{lc} shows that the ratio
$P_{V}/P_{J}$ was slightly less than unity during the 2010 monitoring
period. However, the temporal variations of the PAs were approximately
the same. During the 2005, 2007, and 2009 outbursts, polarization
rotation events were detected for 3C~454.3
\citep{Jorstad10,Sasada10,Sasada12}. However, we did not detect any
polarization rotation events during the 2010 outburst with our frequency
of observations, which was at most once a day.

Figure~\ref{wh} shows the temporal variation of flux in the $V$
band, $\alpha_{VJ}$, the PD in the $V$ band, and the ratio $P_{V}/P_{J}$
for our four-year monitoring period. All the parameters varied,
including the ratio $P_{V}/P_{J}$. During the faint state from
JD~2454800 to 2455050, $\alpha_{VJ}$ became bluer. However, the object
also showed the bluer-when-brighter feature from JD~2454300 to
JD~2454340 \citep{Sasada10,Sasada12}. During the bright states regarded
as outbursts, the parameter $\alpha_{VJ}$ was 1.33$\pm$0.05 (2007;
JD~2454306 to 2454341), 1.34$\pm$0.13 (2008; JD~2454654 to 2454693),
1.48$\pm$0.09 (2009; JD~2455162 to 2455190), and 1.59$\pm$0.11 (2010;
JD~2455502 to 2455532). Here, the error is taken to be the standard
deviation in $\alpha_{VJ}$ The 2010 outburst was the reddest of the four
outbursts.

The PDs in the $V$ band (and also the $J$ band) at the rapid
brightenings of the 2010 outburst were the highest observed during the
four-year monitoring period. The average PDs in the $V$ band during the
outbursts in 2007, 2008, 2009, and 2010 were 7.4\%~$\pm$~3.8\%,
6.1\%~$\pm$~4.6\%, 11.4\%~$\pm$~5.8\%, and 17.4\%~$\pm$~8.1\%,
respectively, where the error is the standard deviation in the PD. The
ratio $P_{V}/P_{J}$ was often less than unity. The average ratios
$P_{V}/P_{J}$ during the 2007, 2008, 2009, and 2010 outbursts were
1.0~$\pm$~0.3, 0.9~$\pm$~0.5, 0.9~$\pm$~0.2, and 0.8~$\pm$~0.1,
respectively. The ratios $P_{V}/P_{J}$ during the 2008, 2009, and
2010 outburst were less than or approximately equal to unity
(Figure~\ref{wh}). Their standard deviations are larger, except in the
2010 outburst for which the ratio $P_{V}/P_{J}$ is less than
unity. Thus, at least FDP occurs during the 2010 outburst.

\section{Discussion}

\subsection{Origin of Frequency-Dependent Polarization}

During the 2010 outburst, 3C~454.3 frequently exhibited FDP along with
the rapid brightenings. We examine two possible causes of the FDP: (1)
contamination from an unpolarized thermal component with spectral
content different than the synchrotron radiation and (2) intrinsic FDP
in the synchrotron radiation.

The object 3C~454.3 was redder-when-brighter during the faint state
(Figure~\ref{wh}), which may be explained by the combination of two
emission processes: thermal radiation from the disk and synchrotron
radiation from the jet. Thermal radiation is less variable and bluer
than synchrotron radiation. In the optical and UV bands, thermal
radiation from an accretion disk is expected to be affected by the
spectral shape \citep{Sasada10,Sasada12}. \citet{Raiteri08b} and
\citet{Sasada10} proposed that the redder-when-brighter trend in
3C~454.3 is because of a substantial contribution of thermal emission
from a disk. When the thermal emission is negligible, this trend shows
``saturation'' at $R{\sim}$14. We estimate the $V$-band saturation
level as $V{\sim}$14.55 by using $B-R=1.2$ and $R{\sim}14$
\citep{Raiteri08b} and assuming that the spectral shape from the $B$ to
$R$ bands follows a simple power-law. The saturation level in the $V$
band is shown as a dotted line in the top panel of Figure~\ref{wh}. Note
that this saturation level is not the flux of the thermal radiation but
the flux level at which the thermal radiation becomes negligible
compared with the observed flux.

In the faint state, the thermal component, being comparable to the
synchrotron component, emerges below the saturation level
\citep{Raiteri08b}. The unpolarized thermal radiation would lower the PD
in the $V$ band compared with that in the $J$ band, because the thermal
radiation becomes more dominant toward the bluer band. For this reason,
the ration $P_{V}/P_{J}$ was less than unity during the faint state from
JD~2454800 to JD~2455000. However, during the rapid brightenings of the
2010 outburst, the object also apparently became  redder and exhibited
FDP, when the peaks were brighter than the saturation level shown in the
top panel of Figure~\ref{wh}. This indicates that the synchrotron
radiation dominated the thermal radiation during the rapid
brightenings. Thus, the FDP was not caused by contamination from thermal
radiation, but was intrinsic to the synchrotron radiation.

\subsection{Frequency-Dependent Polarization of Synchrotron Radiation from Shock Regions}

We will discuss a possible cause of FDP at the synchrotron radiation. 
For electrons with a single power-law distribution of energies, a
theoretical maximum PD of synchrotron emission ($\Pi_{S}$) is; 
\begin{eqnarray}
\Pi_{S} = \frac{p+1}{p+7/3}, \label{pol_Rybicki}
\end{eqnarray}
where $p$ is the power-law index for electron energy distribution
\citep{Rybicki79}. The PD is independent of wavelength, if $p$ is
uniform. However, the polarization in the rapid brightenings exhibited
FDP.

Synchrotron radiation and its polarization in a blazar is caused by
the shock propagating down a relativistic jet with a
turbulent magnetic field \citep{Marscher85,Hughes85}. There could be
multiple shocks and emitting regions even in a single
jet. \citet{Valtaoja88} suggest that frequency-dependent optical
polarization events can be explained with two-component models,
typically consisting of a steep component and a flatter (more highly
polarized) component. If a shock occurs in a jet, particles in the shock
region should be accelerated by a first-order Fermi acceleration
process. High energy particles are generated, and they emit higher
frequency synchrotron radiation. The shock grows rapidly, and the
synchrotron peak moves steeply upward through the infrared/millimeter
regime \citep{Valtaoja91}. Thus, FDP in the optical range could be
caused by the early stages of such shocks. If two synchrotron emissions
exist with different polarization vectors and spectral shapes, the
observed polarization is the sum of the two polarization
vectors. The fractions of these two components in the $V$ and $J$ bands
differ depending on its spectral shape. The PD observed in each band
should also differ, and thus, the object exhibits the FDP. Therefore,
more than two polarization components are required to explain the
observed FDP only by the synchrotron radiation.

\subsection{Shocked Jet Model Applied to Rapid Brightenings}

\subsubsection{Polarization in Shocked Jet Model}


The PD of the synchrotron radiation in blazars depends not only on
the power-law index $p$ of the electron energy distribution but also on
the degree of the alignment of the magnetic field in the shock
region. The PD should be less than the theoretical maximum PD $\Pi_{S}$,
because the magnetic field in the emitting region is turbulent. When the
magnetic field is compressed by the shock, it should be aligned
perpendicular to the direction in which the front of region of shock
moves \citep{Laing80}. In this case, the PD of the synchrotron radiation
from the shock region becomes greater and approaches $\Pi_{S}$. We shall
derive an expression for PD caused by an initially random magnetic field
that is compressed and viewed at an angle to the plane of compression
\citep{Hughes85}. The increase in PD generated by the compression of the
emitting region and the alignment of magnetic field can be represented
as;
\begin{eqnarray}
P\; \approx\; \Pi_{S} \; \frac{(1-{\eta}^{-2})\;{\rm sin}^{2}{\Psi}}{2-(1-{\eta}^{-2})\;{\rm sin}^{2}{\Psi}}, \label{P_Hughes}
\end{eqnarray}
where $\eta$ is the ratio of densities $n$ in the shock region with
respect to the unshocked region, ${\eta}=n_{\rm shock}/n_{\rm unshock}$,
$\Psi$ is the angle between the direction in which the shock moves and
the line of sight in the co-moving frame \citep{Hughes91}. The
inclination angle $\Phi$ of the shock with respect to the jet axis in
the observer's frame is represented by the bulk Lorentz factor $\Gamma$ of the emitting region and the viewing angle $\Psi$ in the co-moving frame: 
\begin{eqnarray}
{\Psi} =
{\rm tan}^{-1}\left[\frac{{\rm sin}\;{\Phi}}{{\Gamma}({\rm cos}\;{\Phi}-\sqrt{1-{\Gamma}^{-2}})}\right]. \label{A_Hughes}
\end{eqnarray}
If this compression by the shock is described by a fluid approximation,
then it is described by the Rankine-Hugoniot equations. In these
equations, $\eta$ is $\eta=(\gamma+1){M}^{2}/[(\gamma-1)M^{2}+2]$,
where $M$ is the Mach number and $\gamma=4/3$ is the specific heat ratio
for the case in which the relativistic jet is constructed from monatomic
molecules. In this case, $\eta$ should be less than 7. For
$\Psi=90^{\circ}$, we can also estimate a lower limit of $\eta$ by using
equation~(\ref{P_Hughes}) with the ratio between the PD
and $\Pi_{S}$.

\begin{figure}
\begin{center}
\begin{tabular}{c}
\includegraphics[scale=0.8]{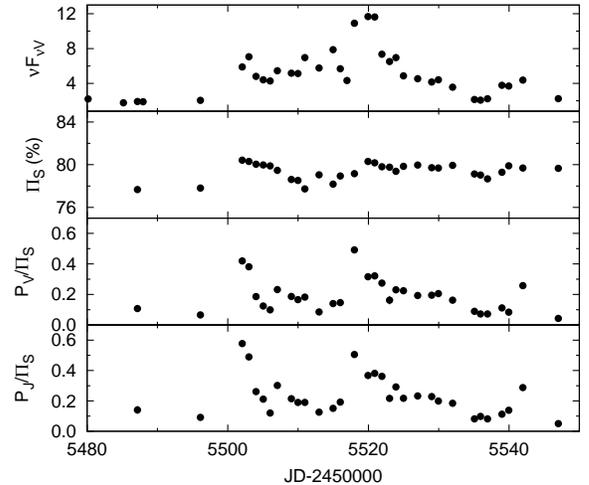}
\end{tabular}
  \caption{Light curve of 3C~454.3 in $V$ band, theoretical
 maximum PD ($\Pi_{S}$) calculated from the observed $\alpha_{VJ}$ and
 the ratios $P_{V}/\Pi_{S}$ and $P_{J}/\Pi_{S}$ between the observed PDs
 and $\Pi_{S}$ in $V$ and $J$ bands. Top panel shows the $V$-band light
 curve during the outburst. A $V$-band flux unit is the same as for
 Figure~\ref{lc}. Second panel shows the temporal variation of $\Pi_{S}$
 (\%) estimated from the observed $\alpha_{VJ}$. Third and bottom panels
 show $P_{V}/\Pi_{S}$ and $P_{J}/\Pi_{S}$, respectively.}
\label{pi}
\end{center} 
\end{figure}

Between the optical and NIR wavelengths, the energy of synchrotron
radiation particles should be approximately similar, because the
$\alpha_{VJ}$ observed in the optical and NIR bands are almost the
same \citep{Sasada10,Sasada12}. In this case, we assume that the
observed $\alpha_{VJ}$ is the spectral index $\alpha$ of synchrotron
radiation and follows $p=2{\alpha}+1$. In the optical and NIR bands, the
$\Pi_{S}$ for synchrotron radiation estimated from $\alpha_{VJ}$ should
be the same. Hereafter, we assume the $\Pi_{S}$ for the optical and NIR
bands are the same. Figure~\ref{pi} shows the light curve in the $V$
band, the temporal variation of $\Pi_{S}$ calculated from
equation~(\ref{pol_Rybicki}) by using the observed $\alpha_{VJ}$, and
the ratio $P_{V}/\Pi_{S}$ and $P_{J}/\Pi_{S}$. The $\Pi_{S}$ were
distributed around 79~\% and the amplitude was less than 4~\%. The
$\Pi_{S}$ is much less variable than the observed PD, and thus the temporal
variation of $P_{V}/\Pi_{S}$ and $P_{J}/\Pi_{S}$ follow the variations
of the observed PDs. \citet{Hughes85} suggest that the power-law index
of the electron distribution does not play a major role in determining
PD. Our result is consistent with this suggestion. 

\subsubsection{Decomposition of Radiations and Application Shocked Jet Model}

Short- and long-term variations both in the flux and polarization
have been reported for blazars 
\citep{Hagen-Thorn08,Villforth10,Sasada11,Sakimoto13,Sorcia13}. Temporal
variations in observed flux and polarization occur on various
timescales. We can assume that the observed synchrotron radiation
consists of short and long-term variable components having different
spectra and different polarization vectors \citep{Sasada11}. In this
case, the observed flux and polarization vector ($F_{\rm obs}$ and 
${\bf P}_{\rm obs}$) are; 
\begin{eqnarray}
F_{\rm obs} = F_{\rm l} + F_{\rm s} \\
{\bf P}_{\rm obs} = {\bf P}_{\rm l} + {\bf P}_{\rm s},
\end{eqnarray}
where fluxes ($F_{\rm l}$ and $F_{\rm s}$) and polarization vectors
(${\bf P}_{\rm l}$ and ${\bf P}_{\rm s}$) are the long- and
short-term components.


The short-term variation is produced by the acceleration at one shock
region in the shocked jet model. Therefore, we can apply
equations~(\ref{P_Hughes}) and (\ref{A_Hughes}) to the short-term
variable components. 
Note that it is important to define
timescales we are looking at, because there are various variation
timescales as can be seen from our four-year monitoring lightcurve
(figure~\ref{wh}). As a result, the behavior of the polarization and
physical parameters estimated from its polarization depend on the choice
of timescales. 3C~454.3 clearly has two
different-timescale variations (timescales of days and a month) for the
2010 outburst (figure~\ref{lc}). Then, the days-timescale rapid
brightenings and month-timescale slowly varing outburst can be regarded
as the short- and long-term components.
We separate $F_{\rm s}$ and ${\bf P}_{\rm s}$ from $F_{\rm obs}$ and
${\bf P}_{\rm obs}$ for the rapid brightenings of the 2010 outburst
using the difference of the variation timescales
and estimate the bulk Lorentz
factors and inclination angles from equations~(\ref{P_Hughes}) and
(\ref{A_Hughes}).

\begin{figure}
\begin{center}
\begin{tabular}{c}
\includegraphics[scale=1]{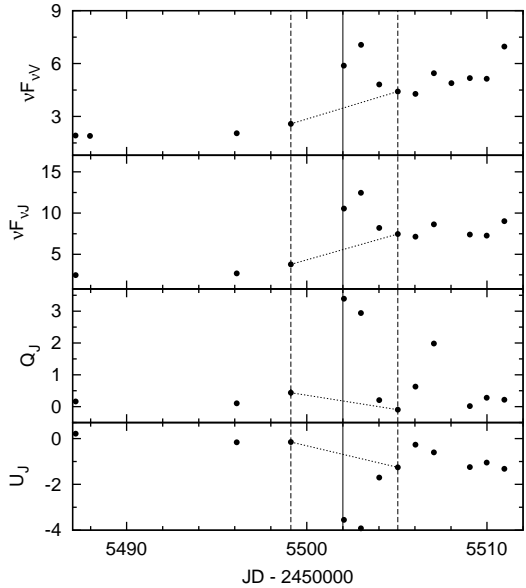}
\end{tabular}
  \caption{Temporal variations around first rapid brightening. From
 top to bottom panels show $V$- and $J$-band light curves and temporal
 variations of Stokes parameters $Q$ and $U$ in the $J$ band. The units
 are {\ergscm}. Solid and dashed lines show the date of the brightenings
 and their pre- and postdates. Dotted line shows the linear
 approximation calculated from the data for pre- and post-brightening
 dates.}
\label{br1}
\end{center} 
\end{figure}

\begin{figure}
\begin{center}
\begin{tabular}{c}
\includegraphics[scale=1]{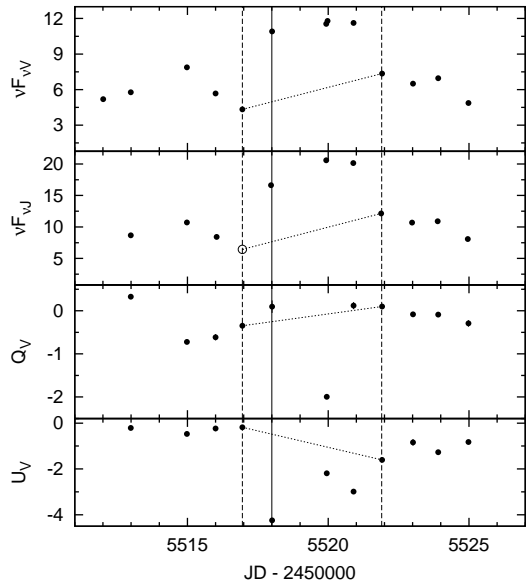}
\end{tabular}
  \caption{Temporal variations around second rapid brightening. From top
 to bottom, panels show a $V$- and $J$-band light curves and temporal
 variations of Stokes parameters $Q$ and $U$ in the $V$ band. Solid,
 dashed, and dotted lines are the same as for Figure~\ref{br1}. Open
 circle in the $J$-band light curve is estimated from the $V$-band flux
 with the assumption that $\alpha_{VJ}=$1.43. See text for details.}
\label{br2}
\end{center} 
\end{figure}

The timescale of the short-term variation is shorter than that of the
long-term variation. We estimate the long-term components in the rapid
brightenings from the pre- and post-brightening data, because the
long-term variation can be regarded as a dominant component in the pre and
postbrightening. We interpolate $F_{\rm l}$ and ${\bf P}_{\rm l}$ at
the brightening by using a linear approximation based on the pre- and
post-brightening data. Figures~\ref{br1} and \ref{br2} show the dates of
the rapid brightenings (solid line) and the pre and postbrightenings
(dashed line). Unfortunately, the data from the predate of the second
brightening was obtained only in the $V$ band, because of the
aforementioned readout error of our instrument. To estimate
$\alpha_{VJ}$, we interpolate the $J$-band flux from the $V$-band flux
by assuming $\alpha_{VJ}$ at the predate of the second brightening to be
the observed averaged $\alpha_{VJ}=1.43$ (JD~24555111 to JD~2455515). We
subtract the estimated $F_{\rm l}$ and ${\bf P}_{\rm l}$ from 
$F_{\rm obs}$ and ${\bf P}_{\rm obs}$ and calculate $F_{\rm s}$ and 
${\bf P}_{\rm s}$. Currently, calculations of polarization are performed
by using the Stokes parameters $Q$ and $U$ in flux units
({\ergscm}). Next, we calculate the ratio between 
$P_{\rm s}$(=$|{\bf P}_{\rm s}|$) and $\Pi_{S}$ estimated from
$\alpha_{\rm s}=1.92$ and 1.52 at the first and second brightenings
($P_{\rm s}/\Pi_{S}$). We also calculate the $F_{\rm l}$, 
${\bf P}_{\rm l}$, and $P_{\rm l}/\Pi_{S}$ in a similar manner. We show
the resultant parameters of the short- and long-term components at the
first and second brightenings in table~\ref{table:br}. The $P_{\rm s}$
at these brightenings are greater than $P_{\rm l}$ and 
$P_{\rm obs}$. $P_{\rm obs}$ of the rapid brightenings indicate that
the appearance of the highly polarized short-term component causes the
increase of $P_{\rm obs}$.

\begin{deluxetable}{cccccc}
\singlespace
\tablecolumns{10}
\tablecaption{\bf Parameters of each component in rapid
 brightenings \label{table:br}}
\tabletypesize{\footnotesize}
\tablehead{
\colhead{}&Band&\colhead{${\nu}F_{\nu}$}&\colhead{$PF$}&\colhead{$P$}&\colhead{$P/{\Pi}_{S}$} \\
\colhead{(1)}&\colhead{(2)}&\colhead{(3)}&\colhead{(4)}&\colhead{(5)}&\colhead{(6)}}
\startdata
1$^{\rm st}$ short & $J$ & 4.95 & 4.30 & 87 & 1.07 \\
1$^{\rm st}$ long  & $J$ & 5.60 & 0.72 & 13 & 0.16 \\
2$^{\rm nd}$ short & $V$ & 5.93 & 3.78 & 64 & 0.81 \\
2$^{\rm nd}$ long  & $V$ & 4.98 & 0.54 & 11 & 0.14 \\
\enddata
\tablecomments{Column 1 - brightenings, 2 - Using band, 3 - total fluxes
 of each component at the brightenings ($\times$10$^{-11}$ {\ergscm}), 4
 - polarized fluxes ($\times$10$^{-11}$ {\ergscm}), 5 - degree of
 polarization calculated from the total and polarized fluxes (\%), 6 -
 ratio between PDs and the $\Pi_{S}$ estimated from $\alpha_{VJ}$.}
\end{deluxetable}

Although the ratio $P_{\rm s}/\Pi_{S}$ for the first brightenings is
greater than unity, $P/\Pi_{S}$ should be less than unity in
principle. The higher value for $P/\Pi_{S}$ can be caused by a systematic
uncertainty of the long-term component estimated by the interpolation of
the linear approximation from the pre- and post-brightening
data. We estimate an error of the ratio $P_{\rm s}/\Pi_{S}$ and
compare the ratio $P/\Pi_{S}$ of the short-term component to that of 
the observed maximum PD with consideration for these errors. The
error of the ratio $P_{\rm s}/\Pi_{S}$, $\delta (P_{\rm s}/\Pi_{S})$, is
calculated by the errors in the total and polarized fluxes and
$\alpha_{VJ}$ of the short-term component. The uncertainty in
$\alpha_{VJ}$ is negligible with respect to $\delta (P_{\rm s}/\Pi_{S})$
because $\alpha_{VJ}$ is less affective to $\Pi_{S}$. The 
$\delta (P_{\rm s}/\Pi_{S})$ of the first and second rapid brightenings
are approximately equal to 0.14 and 0.11 when we assume systematic
errors of 10\% in the total and polarized fluxes of the short-term
component caused by the estimation of long-term component. The assumed
errors of the short-term component are 10 times greater than those of
the observed total and polarized fluxes. The ratio $P_{\rm s}/\Pi_{S}$
for the first brightening is still close to unity even considering large
systematic uncertainties. In this case, the alignment of the magnetic
field in the emitting region of the short-term component is quite
uniform. The ratio $P_{\rm s}/\Pi_{S}$ of the second brightening is also
greater than that at the observed maximum PD. The lower limit of $\eta$
is calculated from the ratio  $P_{\rm s}/\Pi_{S}$ for the second
brightening when $\Psi$ is equal to 90$^{\circ}$, $3.1<\eta<7$.

\begin{figure}
\begin{center}
\begin{tabular}{c}
\includegraphics[scale=0.8]{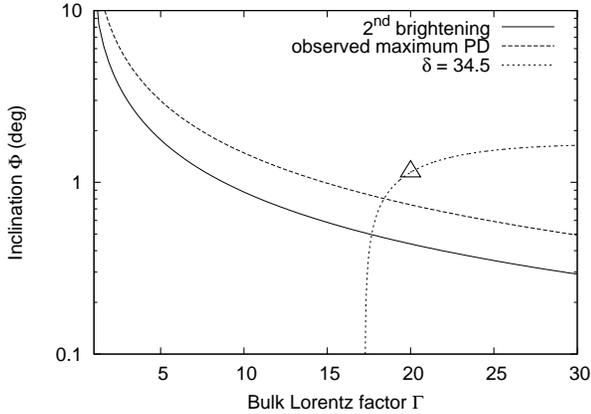}
\end{tabular}
  \caption{Relationship between bulk Lorentz factor $\Gamma$ and
 inclination angle $\Phi$. The solid and dashed lines show the upper
 limit of these parameters estimated from the ratios of PDs and
 equations~(\ref{P_Hughes}) and (\ref{A_Hughes}) (see text for
 details). Solid and dashed lines are calculated from the ratio
 $P_{\rm s}/\Pi_{S}$ for the second brightening and the ratio
 $P_{J}/\Pi_{S}$ from the observed maximum PD. An open triangle shows
 $\Gamma$ and $\Phi$ adopted from \citet{Vercellone11}, and the dotted
 line shows the relationship when the Doppler factor $\delta$ is 34.5.}
\label{para}
\end{center} 
\end{figure}

Using equations~(\ref{P_Hughes}) and (\ref{A_Hughes}), we derived a
constraints on $\Gamma$ and $\Phi$ for the first and second
brightenings. Because $P_{\rm s}$ for the first brightening is close to
$\Pi_{S}$, $\Psi$ should be 90$^{\circ}$ according to
equation~(\ref{P_Hughes}). In this case, $\Phi=1/\Gamma$ as the first
order of approximation. The Doppler factor, 
$\delta=[{\Gamma}(1-{\beta}{\rm cos}\Phi)]^{-1}$ where
$\beta=(1-{\Gamma}^{-2})^{1/2}$, should be equal to $\Gamma$. We can
calculate the limits of the parameters for $\Gamma$ and $\Phi$ by using
equations~(\ref{P_Hughes}) and (\ref{A_Hughes}) and using the ratio
$P/\Pi_{S}$ for the second brightening with the highest case of
$\eta=7$ and at the observed maximum PD for comparison. The curves in
Figure~\ref{para} show the upper limits for $\Gamma$ and $\Phi$. The
values of $\Gamma$ and $\Phi$ should fall below these lines because
$\eta\leq7$. The open triangle shows the parameter for $\Gamma$ and
$\Phi$ adopted by \citet{Vercellone11}, which is estimated from the
multi-wavelength SED from radio to gamma-ray bands around the 2010
outburst. The adopted parameters for $\Gamma$ and $\Phi$ for the outburst
considered by \citet{Vercellone11} is excluded from the acceptable
parameter space estimated from $P_{\rm s}/\Pi_{S}$ and
$P_{J}/\Pi_{S}$. As much as we don't know the origin of the long-term
component, if we apply the shocked jet model to the case
of $P_{\rm l}/\Pi_{S}$ for the first and second brightenings, the
acceptable parameter space of $\Gamma$ and $\Phi$ can allow the
parameter for $\Gamma$ and $\Phi$ adopted by \citet{Vercellone11}.

Although the timescale of the rapid brightenings was several days, the
2010 outburst lasted two months. The rapid brightenings were faster than
the entire variation trend of the outburst. The timescale of variation
becomes faster in proportion to $\delta$. Thus, $\delta$ for the rapid
brightenings can be greater than that of the outburst. In our results,
$\Gamma$ for the first rapid brightening should equal $\delta$ to
explain its high polarization. Thus, $\Gamma$ for the rapid brightenings
can be greater than that of the outburst for these variation timescales.
The emitting region of the rapid brightenings must differ from that of
the outburst to explain why the values of $\Gamma$ for the rapid
brightenings were greater than those of the outburst.

PKS~2155$-$304, a TeV gamma-ray-emitting BL~Lac object, varies on
the timescale from 3 to 5~minutes in the TeV band
\citep{Aharonian07}. \citet{Begelman08} suggest that the bulk Lorentz
factors in the jet must be $\geq$50 to explain this rapid variation. The
Lorentz factor clearly exceeds the jet Lorentz factors $\leq$10. Several
authors propose jet models to explain this complex situation
\citep{Giannios09,Ghisellini09}. Our results indicate that 3C~454.3 is
also involved in two or more components varying on different timescales
with different values for $\Gamma$.

\section{Conclusions}

We monitored blazar 3C~454.3 in the optical and near-infrared bands for
approximately four years starting from July 2007, and we detected two types
of variations in 2010: a large-amplitude outburst occurring on a
timescale of $\sim$months and extraordinary rapid brightenings occurring
on a timescale of $\sim$days. These brightenings had three features: (1)
the $V-J$ bands reddened, (2) the degree of polarization (PD) increased
in the $V$ and $J$ bands, and (3) the polarization was frequency
dependent (FDP). 

Based on these results, we suggest that the observed polarization
vectors can be decomposed into two components, namely long- and
short-term components. We separate the short-term component, estimating
the long-term component for the rapid brightenings from the pre- and
post-brightening data. The estimated PD of the short-term component for
the rapid brightenings was higher than the observed maximum PD and was
close to the theoretical maximum PD. This result indicates that the
short-term emitting regions responsible for the rapid brightenings are
different from the long-term emitting region that radiated the
outburst. The blazar FDP gives us important constraints on $\Gamma$,
$\Phi$, and $\delta$.
\\
\\
This work was supported by a Grant-in-Aid for JSPS Fellows.


\end{document}